\documentclass[twocolumn,aps,prb,10pt,longbibliography]{revtex4-2}
\usepackage{amsmath,amssymb}
\usepackage{bm}
\usepackage{amsthm}
\usepackage{mathrsfs}
\usepackage{multirow}
\usepackage{color}
\usepackage{graphicx}
\usepackage{physics}
\usepackage[colorlinks=true,
            linkcolor=blue,
            citecolor=blue,
            urlcolor=blue]{hyperref}
\usepackage{cancel}

\begin{document}

\title{Quantum Coherence Reshapes Thermodynamic Bounds for Thermal Machines}

\author{Sergi Vidal, Alba Mayor-Fernández and Rosa López\\
Instituto de Física Interdisciplinar y Sistemas Complejos, IFISC (CSIC-UIB)\\
Campus Universitat de les Illes Balears, E-07122 Palma de Mallorca, Spain}

\begin{abstract}
Thermodynamic Uncertainty Relations (TURs) set universal bounds linking current fluctuations to entropy production in nonequilibrium steady states. Their multidimensional generalization (MTUR) introduces matrix inequalities connecting current covariances and mean values. We analyze these bounds in a paradigmatic quantum thermal device, a two-terminal conductor, operating as a heat engine, refrigerator, or heat pump. We show that classical performance limits on efficiency and coefficient of performance remain constrained by the TUR when finite power or heat flow from cold to hot reservoirs is maintained, even in regimes dominated by coherent transport. We further identify the conditions that optimize TUR and MTUR violations, demonstrating that cross-correlations can enhance the joint precision of charge and heat currents near the linear-response regime.
\end{abstract}

\maketitle

\section{\label{sec:intro}Introduction}

The development of microscopic thermal machines capable of converting heat into
work, or vice versa, has emerged as a central challenge in quantum thermodynamics
and nanoscale energy conversion~\cite{Kosloff2013,Vinjanampathy2016,Binder2018,QuantumEnginesReview2024}.
Unlike their macroscopic counterparts, quantum thermal machines operate at length
scales where quantum coherence, thermal fluctuations, and shot noise are all
significant, leading to fundamentally new operating principles and efficiency
limitations~\cite{Scovil1959,Kosloff2013}. Experimental realizations of quantum
thermal machines have been achieved across a wide variety of platforms,
demonstrating that thermodynamic cycles can be operated at the nanoscale.
These include colloidal particles in optical traps implementing
Stirling~\cite{Blickle2012} and Carnot~\cite{Martinez2016} cycles, single trapped
ions~\cite{Abah2012,Rossnagel2014,Rossnagel2016}, nuclear spins~\cite{Peterson2019,Shende2024},
and most recently superconducting transmon qubits acting as the working medium
of a cyclic quantum heat engine~\cite{Uusnaakki2025}. 
In parallel, the theoretical framework of quantum thermodynamics has grown
enormously in recent decades with the overarching goal of understanding the
fundamental bounds on energy conversion at the quantum
scale~\cite{Vinjanampathy2016,Benenti2017}.
\begin{figure}[t]
\raggedright
\includegraphics[width=0.84\columnwidth,clip]{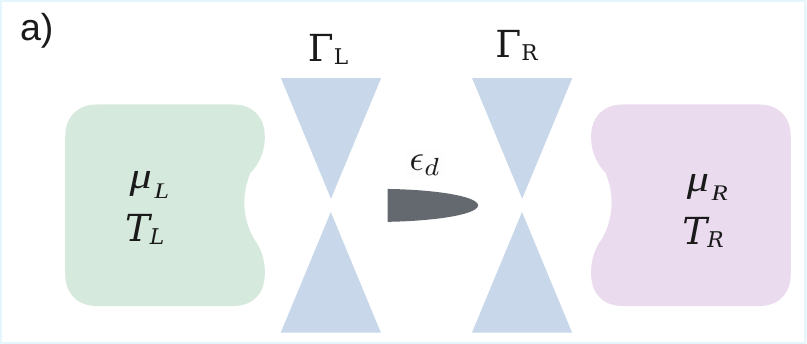}
\includegraphics[width=0.90\columnwidth,clip]{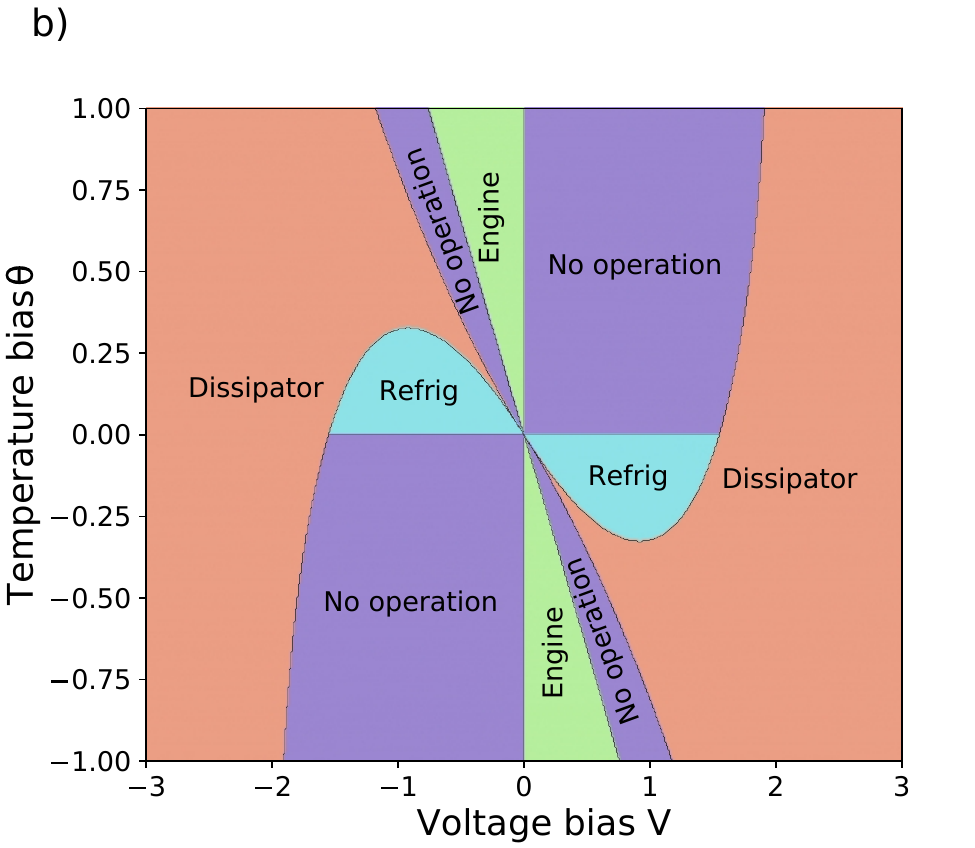}
\caption{a) Schematic representation of the two-terminal coherent thermal machine considered in this work. A single-level conductor with energy $\epsilon_d$ is tunnel coupled to left and right reservoirs characterized by chemical potentials $\mu_L$, $\mu_R$ and temperatures $T_L$, $T_R$, respectively, and is driven out of equilibrium by a voltage bias $V$ and a temperature bias $\theta$ that determine the operating mode (heat engine, refrigerator, or heat pump). b) Operating modes of the system in the ($V$, $\theta$) plane with parameters: $T_0=1$, $\Gamma_{L,R}/T_0 = \Gamma/T_0 = 0.5$, $\epsilon_d/T_0 = 1$. }
\label{fig:1}
\end{figure}

Among the most promising solid-state candidates for quantum thermal machines are
quantum dots~\cite{Josefsson2018,Thierschmann2015,Sothmann2015}. Their discrete
energy spectrum acts as an energy filter that selectively allows electrons to
tunnel only within a narrow energy window, making them natural building blocks
for both thermoelectric conversion and heat management at the
nanoscale~\cite{Staring1993,Benenti2017,Whitney2018}. 
Early theoretical proposals~\cite{Humphrey2002,Edwards1993} established that energy-filtering devices can reach efficiencies close to the Carnot limit, avoiding the need to operate in cycles, with no moving parts ~\cite{Sanchez2011,Jordan2013}. 

These devices have been demonstrated to operate in
distinct thermodynamic modes~\cite{Scovil1959,Kosloff2013,Benenti2017,Pyurbeeva2025}:
as heat engines (converting heat flow into electrical work), as refrigerators
(extracting heat from a cold reservoir using electrical work), or as heat pumps
(transferring heat to a hot side). The experimental realization of these three
modes has been demonstrated across a range of solid-state
platforms~\cite{Josefsson2018,Josefsson2019,Thierschmann2015,Hartmann2015,Prance2009}.
In particular, the experiment by Josefsson \textit{et al.}~\cite{Josefsson2018}
realized a quantum dot heat engine embedded in a semiconductor nanowire and
showed an efficiency exceeding $0.7\,\eta_C$ at finite power, in which both
charge and heat currents were simultaneously measured. 

However, the utility of quantum thermal machines is very much  constrained by fundamental
thermodynamic bounds. The thermodynamic uncertainty relation (TUR), established
by Barato and Seifert~\cite{Barato2015}, provides a universal bound linking
entropy production $\sigma$ to precision  in non-equilibrium steady states.  This relation establishes that
reducing entropy production (approaching reversibility) necessarily requires
reducing the magnitude of mean currents or accepting larger relative fluctuations,
imposing a fundamental precision--dissipation
trade-off~\cite{Shiraishi2016,Maes2017,Horowitz2020,Barato2015, Gingrich2016, Pietzonka2016,Koyuk2020,Dechant2021PRX,Liu2020,Pietzonka2022,Hasegawa2020, Miller2021,VanVu2022,Barato2019,Chun2019,Dechant2018,Dechant2019,Dechant2020PNAS,Dechant2021PRR,Gupta2020,Hiura2021,Koyuk2019,Manikandan2018,Park2021,Pietzonka2017,Proesmans2017,Wolpert2020,Kalaee2021,Brandner2018PRL,Macieszczak2018,Agarwalla2018PRB,Menczel2021,Potts2019,Saryal2019}. This has profound implications for thermal machine design. The simultaneous optimization of efficiency, power, and current stability thus emerges as a central design challenge for nanoscale thermal machines \cite{Benenti2017,Whitney2014}.

In coherent quantum conductors, the classical TUR can be violated,
a property that arises from the quantum-mechanical nature of transport and has
no classical analog~\cite{Brandner2018,Agarwalla2018,Hasegawa2020,Miller2021,VanVu2022,Kalaee2021, Brandner2018PRL,Macieszczak2018,Agarwalla2018PRB,Menczel2021,Potts2019,Saryal2019}. 
Recently, thermodynamic bounds on precision in ballistic multiterminal transport have been derived within the Landauer–Büttiker scattering formalism, showing that TUR violations are generic in the coherent regime and
are ultimately bounded by a modified quantum thermodynamic uncertainty relation \cite{Brandner2025}. These quantum
violations have been studied in quantum dot thermoelectric
junctions~\cite{Liu2019,Agarwalla2018,Brandner2018PRL}, establishing that the interplay between
coherence and dissipation fundamentally reshapes the precision--efficiency
landscape compared to classical stochastic machines.

A natural and powerful generalization of the scalar TUR is the multidimensional
thermodynamic uncertainty relation (MTUR)~\cite{Dechant2019}, which replaces the
single-current bound with a matrix inequality involving the full
covariance matrix of a vector of thermodynamic currents. This formulation captures
cross-correlations between different currents and yields tighter bounds than any
scalar TUR applied independently to each current component. The MTUR is
particularly relevant for quantum thermal machines, where charge and heat currents
are inherently coupled by the energy-filtering mechanism of the quantum dot.

Despite the generality of the TUR and MTUR, the question of whether classical
bounds are violated or saturated differently depending on the operational mode of
a quantum thermal machine remains largely unexplored. Understanding whether engine
or refrigerator operation exhibits fundamentally different trade-offs between
efficiency, power, and noise precision is crucial for optimizing quantum thermal
devices for practical applications~\cite{Josefsson2018,Pyurbeeva2025}.
This question is further motivated by recent advances in the optimal control of
quantum thermal machines~\cite{Erdman2023}, where reinforcement learning has been
used to identify Pareto-optimal power--efficiency trade-off cycles, and by
proposals for novel platforms such as single-molecule heat
engines~\cite{Duan2025} and optically controlled quantum dot
machines~\cite{Liu2020}, which extend the paradigm to new experimental settings.

In this work, we investigate how the TUR and MTUR constrain quantum dot thermal
machines operating in distinct modes: heat engine and refrigerator/heat pump.
We model the system as a single-level quantum dot coupled to two electronic
reservoirs within the Landauer--B\"{u}ttiker scattering formalism, and we explore
the degree to which each bound is saturated in each operational mode, quantified
by saturation parameters $F^\mathrm{x}_{\mathrm{TUR}}$ and
$F_{\mathrm{MTUR}}$ that measure the proximity of the system to the
equality condition in the TUR and MTUR, respectively.
Our key findings are: (i) for engine operation, the TUR is most closely saturated
when the machine prioritizes heat current precision; (ii) for refrigerator/pump
operation, the TUR approaches saturation through charge current precision; (iii) the MTUR saturation parameter $F_\text{MTUR}$ reaches its minimum values symmetrically near the boundary between engine and refrigerator operation, i.e., close to the linear-response regime, suggesting a deep near-equilibrium symmetry between engine and refrigerator operation at the quantum level.

These results suggest a novel perspective on quantum thermal machines: the utility
of coherent two-terminal quantum dots is not simply limited by efficiency rather,
it is stabilized by the interplay between electrical and thermal current
fluctuations. Systems operating near saturation of the TUR bounds experience
intrinsic stability from one current mode while potentially sacrificing precision
in the other. This trade-off, we argue, is intrinsic to quantum mechanical
transport and reflects the fundamental incompatibility between carrying large
currents with low noise in a quantum system.

The paper is organized as follows. Section~\ref{sec:framework} introduces the
theoretical framework within the Landauer--B\"{u}ttiker scattering formalism,
providing expressions for charge and heat currents, current fluctuations, and
their correlation matrix. Section~\ref{sec:results} presents our main results:
the analysis of TUR saturation in different operational modes, the identification
of mode-dependent saturation parameters, the universal behavior revealed by the
MTUR, and the connection between TUR saturation and thermodynamic efficiency.
Section~\ref{sec:conclusions} summarizes our main findings.
\\

\section{Physical Framework}
\label{sec:framework}
We consider a single-level quantum dot coupled to two electronic reservoirs and describe transport within the Landauer--Büttiker scattering formalism \cite{PhysRevB.46.12485}. Within this framework, average charge and heat currents determine the thermodynamic function of the device (engine, refrigerator/pump, or dissipator), while current fluctuations and cross-correlations quantify the stability and precision of such operation. 
Accordingly, the section is structured in two parts. 
Subsection~\ref{sec:sub2A} presents the expressions for charge and heat currents and specifies the corresponding operational modes. 
Subsection~\ref{sec:sub2B} introduces the current fluctuations and their correlation matrix, which serve as the basis for the formulation and evaluation of thermodynamic uncertainty relations.

\subsection{Currents and Operational Modes}
\label{sec:sub2A}

\begin{table*}[t]
\centering
\small
\begin{tabular}{|c|c|c|c|c|c|c|}
\hline
\textbf{Mode} & \textbf{$\dot{W}$} & \textbf{$J_C^Q$} & \textbf{$J_H^Q$} & \textbf{Physical Role} & \textbf{Efficiency} & \textbf{Reversible Limit} \\
\hline\hline
\textbf{Heat engine} & $>0$ & $<0$ & $>0$ & Heat from hot $\to$ Work out & $\eta = \dfrac{|\dot{W}|}{|J_H^Q|}$ & $\eta_C = 1 - \dfrac{T_C}{T_H}$ \\
\hline
\textbf{Refrig./Pump} & $<0$ & $>0$ & $<0$ & Work in $\to$ Extract from cold, deliver to hot & $\text{COP}_{R/P} = \dfrac{|J_{C/H}^Q|}{|\dot{W}|}$ & $\text{COP}_{C,R/P} = \dfrac{T_{C/H}}{T_H - T_C}$ \\
\hline
\textbf{Dissipator} & $<0$ & $<0$ & $<0$ & Work in → Heat dissipated to both reservoirs & --- & --- \\
\hline
\textbf{No-operation} & $< 0$ & $< 0$ & $> 0$ & Work in → Natural heat flow, no useful operation & --- & --- \\
\hline
\end{tabular}
\caption{Operational modes of the two-reservoir quantum dot thermal machine with efficiency definitions and reversible (Carnot) limits. $T_H$ and $T_C$ denote the hot and cold reservoir temperatures, respectively. For the heat engine, the efficiency is defined as the produced work with respect to the heat extracted from the hot reservoir. For refrigerator/heat pump, the distinction is made through the coefficient of performance ($COP_{R/P}$): refrigerator (heat pump) operates to extract (deliver) heat from (to) the cold (hot) side, with external work input requirement ($W < 0$). In the 'No-operation' regime, the electrical bias accelerates the natural heat flow from the hot to the cold reservoir without extracting any useful work or providing cooling. Although work is consumed, this mode is thermodynamically unproductive. The reversible limits represent the theoretical maximum efficiency achievable in each mode, given by the Carnot theorem.}
\label{tab:modes}
\end{table*}

In the present work we consider a two-terminal setup, with left ($L$) and right ($R$) reservoirs characterized by chemical potentials $\mu_\alpha$ and temperatures $T_\alpha$ ($\alpha=L,R$). 
Each reservoir is described by the Fermi--Dirac distribution
\begin{equation}
f_\alpha(\epsilon) = \frac{1}{1 + \exp[(\epsilon - \mu_\alpha)/(k_B T_\alpha)]}. 
\end{equation}
We introduce the average temperature and chemical potential,
\(
T_0=(T_L+T_R)/2
\)
and
\(
E_F=(\mu_L+\mu_R)/2
\),
and assume symmetric electrical and thermal biases around these values,
\begin{equation}
T_{L,R}=T_0\pm\frac{\theta}{2}, 
\qquad
\mu_{L,R}=E_F\pm\frac{eV}{2}.
\end{equation}
The quantum dot acts as a resonant scatterer with a single level at energy $\epsilon_d$. 
Its transmission probability takes the Breit--Wigner (Lorentzian) form
\begin{equation}
T(\epsilon) = 
\frac{\Gamma_L\Gamma_R}
{(\epsilon - \epsilon_d)^2 + (\Gamma_L + \Gamma_R)^2/4},
\end{equation}
where $\Gamma_{\alpha\in \{L,R\}}$ denotes the tunneling coupling to reservoir $\alpha$ that we consider constant within the wide band limit approach. For future considerations $\Gamma=\Gamma_L =\Gamma_R$.
The charge current from lead $L$ to the quantum dot is given by the Landauer formula:
\begin{equation}
I_L = \frac{e}{h} 
\int d\epsilon \,
T(\epsilon)
\left[ f_L(\epsilon) - f_R(\epsilon) \right],
\label{eq:current_landauer}
\end{equation}
The right-lead current $I_R$ follows analogously with $L \leftrightarrow R$, and charge conservation gives $I_L + I_R = 0$ in steady state.
The heat current from lead $L$ is:
\begin{equation}
J_L^Q = \frac{1}{h} 
\int d\epsilon \,
(\epsilon - \mu_L) \,
T(\epsilon)
\left[ f_L(\epsilon) - f_R(\epsilon) \right].
\label{eq:heat_current}
\end{equation}
The energy conservation law is translated into
\begin{equation}
\sum_{\alpha} J^Q_\alpha - \dot{W} = 0,
\end{equation}
where the electrical work power is (using $I_R = -I_L$)
\begin{equation}
\dot{W}= -\sum_\alpha \mu_\alpha I_\alpha = -\mu_L I_L - \mu_R I_R=-\Delta\mu I_L .
\end{equation}
Notice that our sign convention is that currents leaving the electronic reservoirs are considered positive, as well as the work performed by the system.
The two-reservoir QD can operate as a thermal engine for some electro-thermal configurations, i.e. it can generate power, pump heat or refrigerate a cold bath depending on the electrical bias $eV$ and the thermal gradient $\theta$. 
The operational modes are defined based on the useful thermodynamical task that the system is able to perform, which are detailed in Table~\ref{tab:modes}.

The two-terminal QD setup enters the heat engine regime when the heat flow from the hot to the cold reservoir generates electrical work pushing the electron flow against the applied bias voltage. 
The efficiency is defined as 
\begin{equation}
    \eta = \frac{|\dot W|}{|J_H^Q|}
\end{equation} 
For positive $\theta$ it corresponds to the left contact, $J_L^Q$, whereas for the negative $\theta$ it is the right contact, $J_R^Q$.

On the other hand, our setup can perform a refrigeration task by extracting heat from the cold reservoir when it is electrically powered by $eV$. In this case, the coefficient of performance is defined as
\begin{equation}
\text{COP}_R = \frac{|J_C^Q|}{|\dot W|}.
\end{equation}
In the case where our interest is to deliver heat to the hot reservoir, our two-terminal conductor operates as a heat pump. Its coefficient of performance is defined as
\begin{equation}
\mathrm{COP}_P = \frac{|J_H^Q|}{|\dot{W}|},
\end{equation}
i.e., the heat delivered to the hot reservoir per unit of input work.

Notice that our setup can operate without performing any thermodynamical task, but this regime is not of our interest for thermal machine applications.

Throughout this work, unless otherwise stated, we consider $E_F/T_0 = 0$, $\epsilon_d/T_0 = 0.5$, $\Gamma_L/T_0 = \Gamma_R /T_0= 0.5$.
The thermodynamical useful modes, engine or refrigerator/heat pump occur whenever the  particle-hole (electron-hole) symmetry is broken, i.e. when $\epsilon_d \neq  E_F$  ; otherwise the setup is not able to generate thermoelectrical power as we explicitly demonstrated in  Appendix~\ref{app:phsymmetry}. Besides, we have chosen for our simulations intentionally symmetric tunnel couplings, i.e., $\Gamma_L=\Gamma_R$) since this choice simplifies the analysis and, as discussed in Appendix~\ref{app:tursat}, leads to the tightest saturation of the TUR for the electric current. 
For the chosen device's parameter configuration the setup is encountered in any of the four different modes described in Table~\ref{tab:modes}. Here, the different modes in the ($V$, $\theta$) plane are characterized by a color code. 
Fig.\textcolor{blue}{~\ref{fig:1}b)} displays a green triangular zone that corresponds to the engine mode. This domain reflects a monotonic relationship where increasing temperature bias allows progressively larger electrical loads to be overcome, with the slanted boundary marking the electrochemical threshold beyond which power generation ceases. 
In stark contrast, the blue semi-elliptical refrigerator region exhibits non-monotonic behavior. There is an optimal voltage (at the peak of the region) that maximizes heat extraction beyond which further increases in $|V|$ are counterproductive as generates too much dissipation.
This optimum arises from the fundamental competition between electrostatic heat pumping and quadratic Joule heating. At small voltages, work is efficiently converted to cooling, but as $|V|$ increases, resistive dissipation increasingly dominates, eventually reducing net refrigeration. 
The orange dissipative zones represent regimes where the system  does no useful thermodynamic function, the power facilitates the heat flow towards both reservoirs, the hot and the cold ones, dissipating heat. 

Now, once we have shown the four device's modes in our setup we determine in which conditions it performs a thermodynamic useful task together with the requirement of being stable. 
Stability performance is limited by the TUR. In the following we present the calculation of the current fluctuations and their relation with the TUR and the MTUR.
\subsection{Current--current correlations and TURs}
\label{sec:sub2B}
The performance of a multiterminal thermoelectric device is determined not only by mean currents but also by their fluctuations. 
Charge, heat, and mixed charge--heat noises quantify the constancy of operation, and thermodynamic uncertainty relations connect these noises to dissipation.

In particular, for a two-terminal conductor, the autocorrelation function for the $\mathrm{x}$-current  (with $\mathrm{x}\in\{$ charge, heat $\}$) in the left contact
($S\equiv S_{LL}$) reads as
\begin{equation}
\begin{split}
S_\mathrm{x} = \int d\epsilon \, X(\epsilon)\Big[
T(\epsilon)\big(1-T(\epsilon)\big)
\big(f_L(\epsilon)-f_R(\epsilon)\big)^2  \\
+\, T^2(\epsilon)\Big\{
f_L(\epsilon)\big(1-f_L(\epsilon)\big)
+ f_R(\epsilon)\big(1-f_R(\epsilon)\big)
\Big\}
\Big],
\end{split}
\label{eq:noise_compact}
\end{equation}
which naturally separates into shot-noise and thermal-noise contributions,
$S_\mathrm{x}=S_\mathrm{x}^{\mathrm{sh}}+S_\mathrm{x}^{\mathrm{th}}$.
The first term, proportional to $T(1-T)$, describes partition (shot) noise and vanishes in equilibrium.
The second term, proportional to $f(1-f)$, corresponds to thermal noise arising from equilibrium fluctuations in the reservoirs and remains finite even in the absence of applied biases. The prefactor $X(\epsilon)$ specifies the type of current considered.
For charge noise ($\mathrm{x}=I$, with $I\equiv I_L$) one has $X(\epsilon)=e^2/h$;
for heat noise ($\mathrm{x}=J$, with $J\equiv J_L^Q$), $X(\epsilon)=(\epsilon-\mu_L)^2/h$;
and for the mixed charge--heat correlator ($\mathrm{x}=IJ$), $X(\epsilon)=e(\epsilon-\mu_L)/h$. 
The full analytic expression for the thermal contribution to the charge-current autocorrelation is derived in Appendix~\ref{app:thermalnoise}.




%
The complete set of charge and heat fluctuations is collected in the symmetric $2\times 2$ matrix
\begin{equation}
\boldsymbol{\mathcal{S}}
=\begin{pmatrix}
S_I & S_{IJ} \\
S_{IJ} & S_J
\end{pmatrix},
\label{eq:noise_matrix_2terminal}
\end{equation}
which is positive semi-definite, $\det\boldsymbol{\mathcal{S}}\ge 0$.
Its eigenvalues define the principal fluctuation channels: when $S_{IJ}^2 = S_I S_J$ the charge and heat currents are fully correlated and $\boldsymbol{\mathcal{S}}$ has rank one; otherwise they are only partially correlated.
The determinant $\det\boldsymbol{\mathcal{S}}$ appears inverted in thermodynamic uncertainty relations, making small determinants correspond to a large entropy-production cost for low-noise operation. The expression for the TUR reads
\begin{equation}
\frac{\text{Var}(\mathrm{x})}{\langle \mathrm{x} \rangle^2} \geq f(\sigma).
\label{eq:TURs_general_form}
\end{equation}
Initially, thermodynamic uncertainty relations were derived for classical Markovian processes obeying local detailed balance, yielding the bound
$f(\sigma)=2k_B/\sigma$ \cite{Barato2015}.
In coherent quantum conductors, however, this classical bound can be violated, leading to modified quantum thermodynamic uncertainty relations (QTUR) that account for phase-coherent transport \cite{Brandner2025}.

Moreover, other generalizations have been made, as the MTUR, which involves an $n$--dimensional vector of thermodynamic quantities, with $n \geq 1$ \cite{Dechant2019}.  This version  relates several quantities with not only self-correlations, but also with cross correlations, which can yield tighter bounds on the relative current precision, i.e., the signal-to-noise ratio $\mathrm{SNR} \equiv \langle J \rangle^2 / \mathrm{Var}(J)$. For this reason, cross-correlations between different observables play a significant role, such as those between distinct charge or heat currents, as considered in this work. In a general framework, the MTUR for a current vector $\mathbf{J} = (J_1, J_2, \ldots, J_n)^T$ is formulated accordingly to:
\begin{equation}
\sigma \geq 2 k_B \langle \mathbf{J} \rangle^T \mathcal{S}^{-1} \langle \mathbf{J} \rangle.
\label{eq:mtur_general}
\end{equation}

The entropy production rate $\sigma$ is defined through the Clausius relation
\begin{equation}
    \sigma = - \sum_{\alpha}\frac{J^Q_\alpha}{T_\alpha}.
\end{equation}
By the second law of thermodynamics, $\sigma \geq 0$ in all steady-state operational modes, with $\sigma = 0$ only at thermodynamic equilibrium.

A quantitative indicator of how tightly the TUR is saturated or violated  is obtained by defining, for current~$\mathrm{x}$ (charge or heat at contact $\alpha \in \{L,R\}$, with $S_{\rm x}$ the corresponding self-correlator):
\begin{equation}
    F_{\rm TUR}^{\mathrm{x}} = \frac{S_\mathrm{x}}{\mathrm{x}^2}\frac{\sigma}{k_B} - 2.
\end{equation}

A negative value $F^\mathrm{x}_\text{TUR} < 0$ signals a violation of the TUR. Hereafter, our analysis focuses on the charge or heat current at the left contact , i.e., $\mathrm{x}=I_L$ or, $\mathrm{x}=J_L^Q$ to be considered in the $F_{\rm TUR}$.  In a similar fashion, a quantitative indicator of how tightly the MTUR is saturated when the charge and heat flows are jointly considered, with  $\mathbf{J}~=~(I_L, J_L^Q) $, reads  
\begin{equation}
    F_{\rm MTUR} = \frac{\sigma/k_B}{\langle \mathbf{J} \rangle^T \mathcal{S}^{-1} \langle \mathbf{J} \rangle}-2.
\end{equation}
Similarly, $F_{\rm MTUR}<0$ indicates that the MTUR is violated.
For the numerical calculations presented below we set $k_B=e=h=1$.
\section{Results}
\label{sec:results}

In this section we analyze how thermodynamic uncertainty relations constrain the operation of the two-terminal quantum-dot thermal machine across its different modes (heat engine, refrigerator/heat pump, and dissipator). Our goal is to understand how tightly the bounds are saturated in each regime, and what this reveals about the stability and precision of charge and heat transport.

We proceed in three steps. First, in Secs.~\ref{sec:chargetur} and \ref{sec:heattur}, we examine the single-current TUR for charge and heat separately. Next, in Sec.~\ref{sec:mtur}, we apply the MTUR to the coupled charge–heat currents. Finally, in Sec.~\ref{sec:efficiency}, we relate the saturation of the TUR and MTUR to thermodynamic efficiency.

Let us begin by examining the \emph{single-current} TUR applied separately to the charge and heat currents. We compute $F_{\mathrm{TUR}}^{I}$ for the electrical current and $F_{\mathrm{TUR}}^{J}$ for the heat current, and map their behavior over the ($V, \theta$) diagram. Our results show that these indicators behave very differently depending on the operational mode suggesting that, depending on the thermodynamic task, the device selectively stabilizes either the electrical or the thermal sector.
We then turn to the \emph{multidimensional} TUR, which treats charge and heat currents on equal footing through their full covariance matrix. Here, we explore the multidimensional saturation parameter $F_{\mathrm{MTUR}}$ and study its dependence on the electrical and thermal biases. Unlike the single-current indicators, $F_{\mathrm{MTUR}}$ exhibits a more symmetric behavior between engine and refrigerator modes and reaches its minimum values near equilibrium. This indicates that the \emph{joint} precision of charge and heat currents is optimized in a neighborhood of the linear-response regime, where cross-correlations between charge and heat fluctuations play a crucial role. 
\subsection{Charge Thermodynamic Uncertainty Relation}
\label{sec:chargetur}
\begin{figure}[h!]
\raggedleft
\includegraphics[width=0.95\columnwidth,clip]{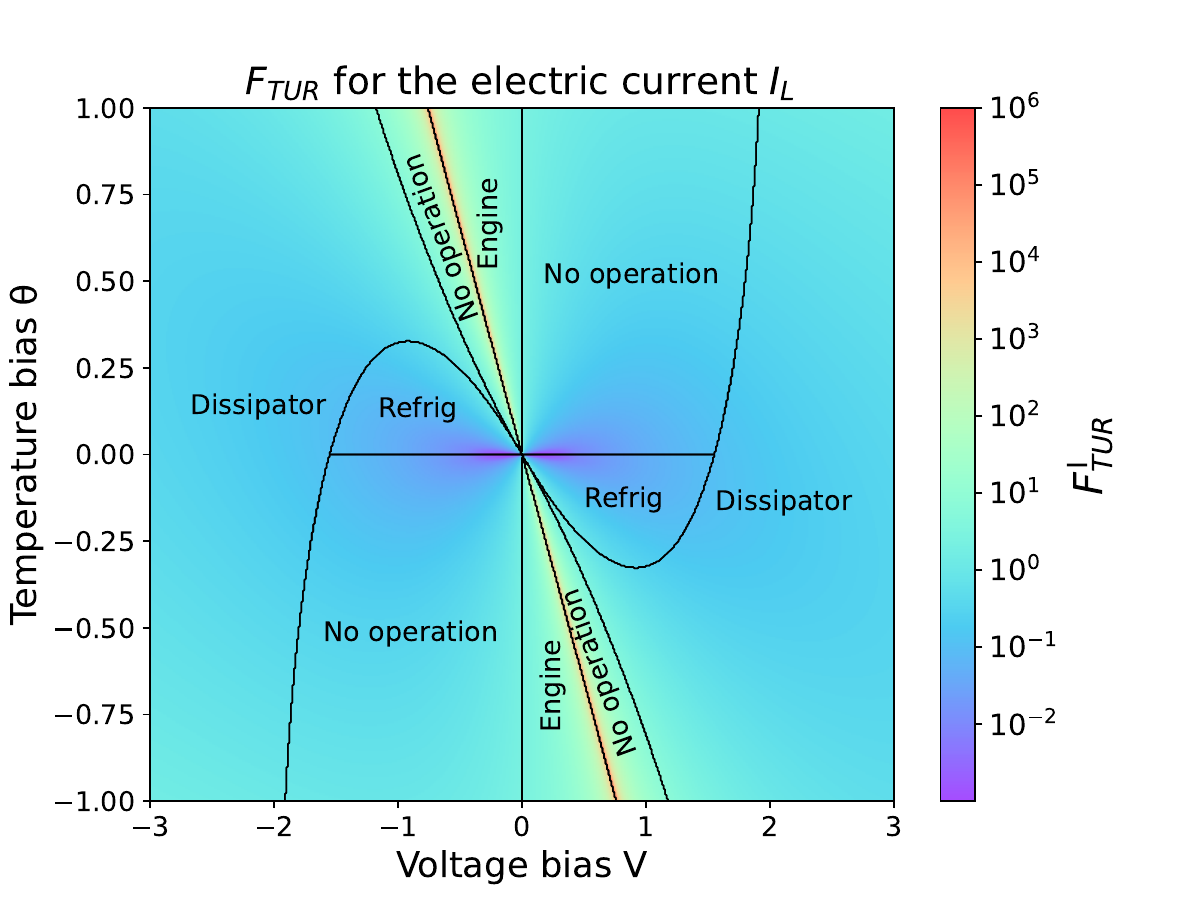}
\caption{Colormap of the TUR factor for the electric current in the left reservoir, $F_{\mathrm{TUR}}^{I}$, as a function of the temperature bias and electric voltage difference between the terminals. For clarity, values of $F_{\mathrm{TUR}}^{I}$ larger than $10^5$ are saturated and displayed with the same color. Machine operation regimes are indicated. Parameters: $\Gamma_{L,R}/T_0 = \Gamma/T_0 = 0.5$, $\epsilon_d/T_0 = 0.5$.}
\label{fig:tur_electricurrent1}
\end{figure}
We start by analyzing Fig.~\ref{fig:tur_electricurrent1} which shows the saturation parameter $F_{\mathrm{TUR}}^{I}$ for the left charge current $I_L$ across the $(V, \theta)$ plane, with the operational modes shown. The plot reveals a clear qualitative trend: the TUR associated with the charge current is \emph{best saturated} (smallest $F_{\mathrm{TUR}}^{I}$) when the device operates as a refrigerator/heat pump, while markedly larger values of $F_{\mathrm{TUR}}^{I}$ appear in the engine regime. We observe that in the refrigerator region, $F_{\mathrm{TUR}}^{I}$ takes values close to zero over a broad parameter range, indicating that the charge current operates close to the TUR equality. Physically, this means that when the dot is used as a cooling device, the electrical current which supplies the work input is nearly \emph{precision-optimal}.  For the observed entropy production, its fluctuations are almost as small as the TUR permits. In other words, the machine uses its thermodynamic budget to stabilize the \emph{input} (driving) current. In sharp contrast, when the same setup is operated as a heat engine, $F_{\mathrm{TUR}}^{I}$ becomes large. The electrical current, now the \emph{useful output} carrying power against the bias, is far from the TUR bound: for a given $\sigma$, its fluctuations are much larger than the minimal TUR value. The engine therefore converts a relatively regular thermal input (as demonstrated in Fig. 3 and Sec.~\ref{sec:heattur}) into a comparatively noisy electrical output. From the perspective of precision, the engine mode is not optimized for charge-current stability.

\subsection{Heat Thermodynamic Uncertainty Relation}
\label{sec:heattur}
\begin{figure}[t]
\raggedleft
\includegraphics[width=0.95\columnwidth,clip]{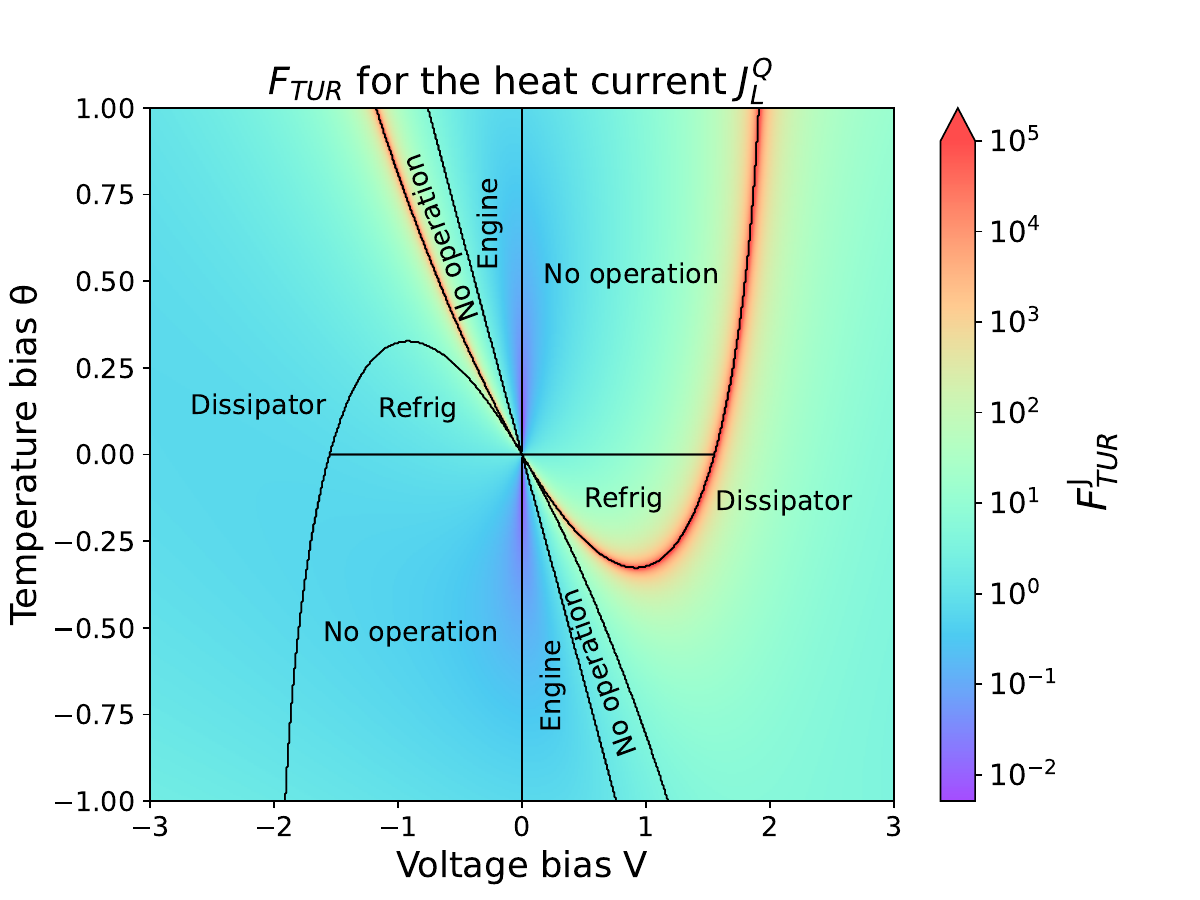}
\caption{Colormap of the heat current TUR factor (left reservoir) $F_{\mathrm{TUR}}^{J}$ as a function of the temperature gradient and voltage bias. For clarity, values of $F_{\mathrm{TUR}}^{J}$ larger than $10^5$ are saturated and displayed with the same color. Machine operation regimes are indicated. Parameters: $\Gamma_{L,R}/T_0 = \Gamma/T_0 = 0.5$, $\epsilon_d/T_0 = 0.5$.}
\label{fig:tur_heatcurrent1}
\end{figure}

Fig.~\ref{fig:tur_heatcurrent1} presents the thermal counterpart of $F^I_\mathrm{TUR}$ shown in Fig. 2, namely the saturation parameter $F^J_\mathrm{TUR}$ for the heat current $J_L^Q$ from the left reservoir. Remarkably, the qualitative behavior is \emph{inverted} compared to Fig.~\ref{fig:tur_electricurrent1}.  Now, in Fig.~\ref{fig:tur_heatcurrent1} the TUR is best saturated in the \emph{engine} regime, while larger values of $F_{\mathrm{TUR}}^{J}$ are found when the dot operates as a refrigerator. In the engine domain, the region of small $F_{\mathrm{TUR}}^{J}$ indicates that the heat flow from the hot reservoir is close to the TUR equality limit. Thus, in engine mode, the \emph{thermal input} $J_L^Q$ is the current that is made as regular as possible given the entropy production. This is precisely the thermodynamic resource that fuels power generation, and the device uses its dissipation budget to stabilize this resource current. Conversely, in the refrigerator region, $F_{\mathrm{TUR}}^{J}$ becomes large. Here, the heat current is now the \emph{useful output} (controlled extraction from or injection into a reservoir), and it carries the larger relative noise. The machine stabilizes the electrical work input at the expense of letting the thermal output fluctuate more.

Taken together, Figs.~\ref{fig:tur_electricurrent1} and \ref{fig:tur_heatcurrent1} reveal a consistent pattern. In each operational mode, the TUR is most tightly saturated for the current that plays the role of \emph{driving resource} (work in the refrigerator, heat in the engine), while the \emph{driven, useful output} current is the one furthest from the TUR bound. This “role reversal” between charge and heat currents is an intrinsic precision trade-off: at fixed entropy production, the machine can operate near the TUR limit for one sector (electrical or thermal), but not for both simultaneously.

\subsection{Joint charge--heat precision and MTUR}
\label{sec:mtur}
\begin{figure}[h!]
\raggedleft
\includegraphics[width=0.95\columnwidth,clip]{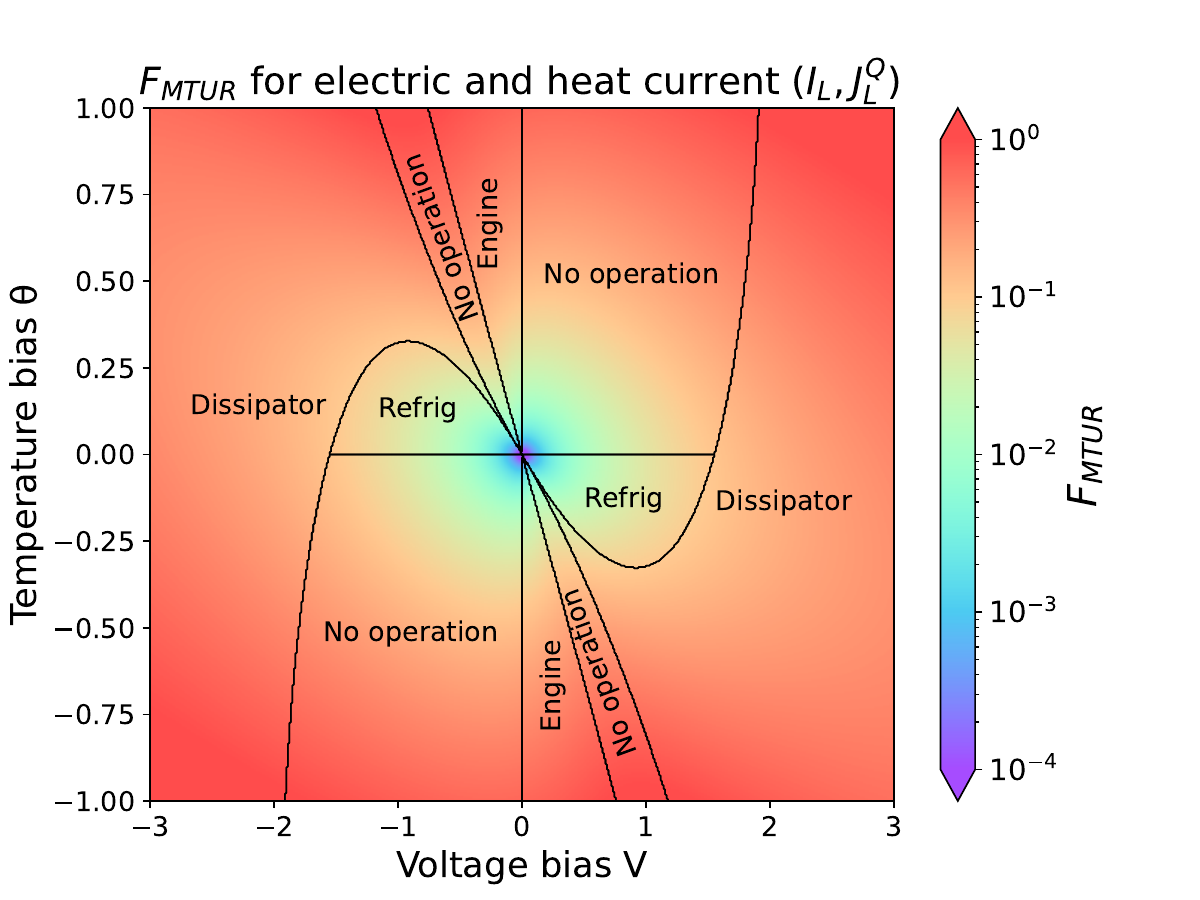}
\caption{Colormap of the MTUR factor, $F_{\mathrm{MTUR}}$,  for the electric and heat currents in the left reservoir $I_L, J^Q_L$, respectively, as a function of the temperature gradient and voltage bias between the two terminals. Machine operation regimes are indicated. For clarity, values of $F_{\mathrm{MTUR}}$ smaller than $10^{-4}$ are saturated and displayed with the same color. Parameters: $\Gamma_{L,R}/T_0 = \Gamma/T_0 = 0.5$, $\epsilon_d/T_0 = 0.5$.}
\label{fig:mtur}
\end{figure}
The MTUR combines charge and heat currents into a single matrix inequality. The associated saturation parameter $F_{\mathrm{MTUR}}$, plotted in Fig.~\ref{fig:mtur}, measures how tightly this \emph{joint} bound is fulfilled when charge and heat currents are treated on equal footing. Several salient features emerge from Fig.~\ref{fig:mtur}, namely, (i) the \emph{minimum} values of $F_{\mathrm{MTUR}}$ are reached in a band that runs near the boundary between the engine and refrigerator regions, i.e., close to the linear-response regime around equilibrium. In this vicinity, $F_{\mathrm{MTUR}}\approx 0$, indicating that the coupled charge--heat currents jointly saturate the MTUR bound. (ii) In contrast to the single-current indicators $F_{\mathrm{TUR}}^{I}$ and $F_{\mathrm{TUR}}^{J}$, the MTUR saturation is comparatively \emph{symmetric} between the engine and refrigerator regimes: both modes exhibit parameter regions where $F_{\mathrm{MTUR}}$ is small. The multidimensional bound does not care whether the machine is primarily stabilizing charge or heat. It constrains the \emph{combination} $(I_L,J_L^Q)$. Finally (iii) as one moves far from equilibrium (large voltage and/or large temperature biases), $F_{\mathrm{MTUR}}$ grows, reflecting the increased entropy-production cost of maintaining simultaneously precise charge and heat flows in strongly driven regimes.

These observations have two key implications. First, they show that while each single current can be made precision-optimal only in one operating mode (charge in the refrigerator, heat in the engine), the \emph{joint} pair $(I_L,J_L^Q)$ can approach the MTUR bound in \emph{both} modes near equilibrium. Second, this highlights the crucial role of cross-correlations $S_{IJ}$.  Near equilibrium, charge and heat fluctuations are strongly correlated in such a way that the combination entering $\mathbf{J}^T \mathcal{S}^{-1} \mathbf{J}$ is better constrained than either variance alone.
\subsection{Correlation with efficiency}
\label{sec:efficiency}
\begin{figure}[h!]
\raggedleft
\includegraphics[width=0.95\columnwidth,clip]{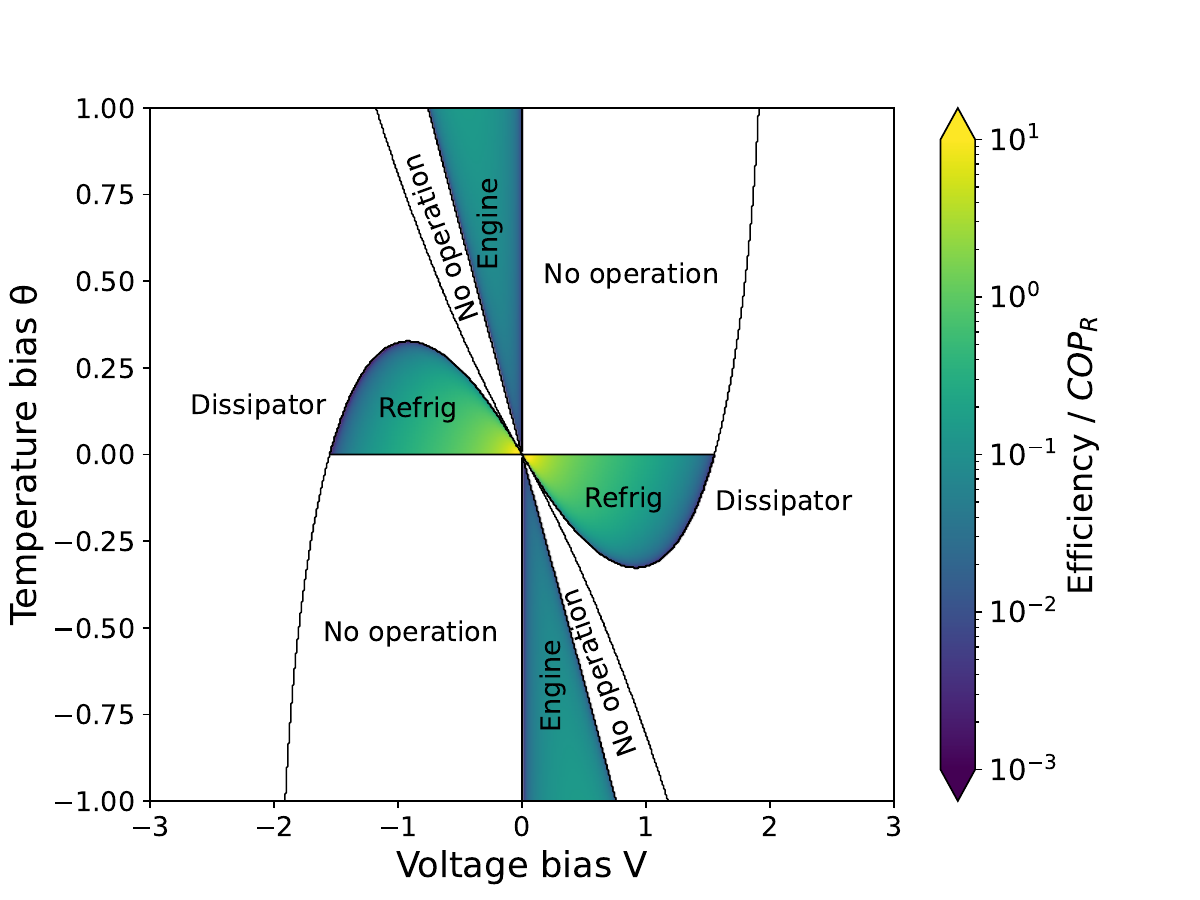}
\caption{Heatmap plot of the efficiencies reached for the different operational modes discussed in Table 1. Parameters: $\Gamma_{L,R}/T_0 = \Gamma/T_0 = 0.5$, $\epsilon_d/T_0 = 0.5$. Values larger than $10$ and smaller than $10^{-3}$  are saturated and shown with the same color. 
}
\label{fig:efficiencies}
\end{figure}
Finally, we connect the observed saturation patterns with the thermodynamic performance of the device in its different operational modes.
Using Fig.~\ref{fig:efficiencies}, we compare the regions where
\(F_{\mathrm{TUR}}^{I}\), \(F_{\mathrm{TUR}}^{J}\), and \(F_{\mathrm{MTUR}}\) are small with the corresponding engine efficiencies and refrigeration coefficients of performance defined in Table~\ref{tab:modes}.
This comparison reveals a structured precision--efficiency trade-off.
Regimes in which an individual current nearly saturates the TUR do not necessarily correspond to maximal efficiency. 
In contrast, regions where the MTUR approaches saturation tend to coincide with enhanced performance, particularly in the refrigeration regime. This correspondence is most pronounced close to equilibrium, where both the MTUR bound becomes tighter and the coefficient of performance approaches its maximal values.
By combining Figs.~\ref{fig:tur_electricurrent1}, \ref{fig:tur_heatcurrent1}, \ref{fig:mtur}, and \ref{fig:efficiencies}, we can directly relate TUR/MTUR saturation to thermodynamic performance.
In the refrigerator regime, where \(F_{\mathrm{TUR}}^{I}\) is small, the coefficient of performance reaches values of order \(10^{0}\!-\!10^{1}\), indicating relatively high cooling efficiency achieved with a remarkably regular input current and moderate entropy production.
In the engine regime, where \(F_{\mathrm{TUR}}^{J}\) is small, the efficiency can reach values of the order of \(\sim 10^{-1}\!-\!10^{0}\): the heat inflow becomes stabilized, while the output power remains comparatively noisy.
Finally, close to equilibrium, where \(F_{\mathrm{MTUR}}\) attains its minimum, the efficiency in the refrigerator region is maximal for the chosen parameter set.
In this regime the device operates in its most thermodynamically balanced configuration: neither charge nor heat currents are individually TUR-optimal, yet their joint fluctuations are constrained as tightly as allowed by the entropy production.
\section{Conclusions}
\label{sec:conclusions}
In this work, we have investigated how thermodynamic uncertainty 
relations in the scattering theory framework both in their scalar (TUR) and multidimensional (MTUR) 
formulations   constrain the operation of a coherent two-terminal quantum dot thermal machine across its distinct functional modes:  heat engine refrigerator/heat pump.

Our central finding is that the TUR and MTUR do not simply impose a uniform scalar ceiling on precision, but instead carve out a structured precision landscape in the joint charge–heat current 
space $(I_L, J_L^Q)$. This landscape is strongly mode-dependent, i.e., the TUR is most tightly saturated for whichever current plays the role of the \emph{driving resource} in each operational mode. In the refrigerator regime, where electrical work is the input resource, the charge current $I_L$ operates closest to the TUR bound, exhibiting near-minimal relative fluctuations for the available entropy production. Conversely, in the heat engine regime, it is the heat current $J_L^Q$ the thermodynamic fuel powering work extraction that becomes precision-stabilized, while the output electrical current is comparatively noisy. This \emph{role reversal} between charge and heat current precision is a fundamental and intrinsic feature of quantum-coherent two-terminal transport, reflecting the incompatibility of simultaneously minimizing fluctuations in both currents at fixed entropy production.
The multidimensional TUR provides a deeper and more symmetric 
perspective on this trade-off. Unlike the scalar TUR, which 
distinguishes sharply between engine and refrigerator operation, 
the MTUR saturation parameter $F_\mathrm{MTUR}$ reaches its minimum values in a band that runs near the boundary between both operational modes, i.e., close to the linear-response regime. 
This near-equilibrium universality reveals that the \emph{joint} 
precision of charge and heat currents captured through their 
full covariance matrix including cross-correlations $S_{IJ}$ 
is optimized symmetrically across both modes. The cross-correlation between charge and heat fluctuations plays a decisive role here: 
near equilibrium, these correlations allow the coupled pair $(I_L, J_L^Q)$ to approach the MTUR bound more tightly than either current alone could achieve independently.

We have also established a direct connection between TUR/MTUR saturation and thermodynamic performance. In the refrigerator regime, the near-saturation of the charge TUR coincides with relatively large coefficients of performance, indicating that high cooling efficiency is achieved at the cost of a stabilized electrical input.

In summary, thermodynamic uncertainty relations, particularly 
in their multidimensional formulation, reveal that coherent 
quantum dot thermal machines operate under a rich and 
structured precision–efficiency landscape. The device does 
not simply trade efficiency for power, it selectively 
stabilizes the current that matters most for its 
thermodynamic task, and the MTUR encodes the deepest 
constraint on this stability, pointing toward a universal 
near-equilibrium precision limit shared by engines and 
refrigerators alike.
\section{Acknowledgments}
We thank Gianmichele Blassi and Sungguen Ryu for fruitful discussions. 
We acknowledge support by the Spanish State Research
Agency (MCIN/AEI/10.13039/501100011033) and FEDER
(UE) under Grant No. PID2020-117347GB-I00 and María
de Maeztu Project No. CEX2021-001164-M.

%
%

\bibliographystyle{apsrev4-2}
\bibliography{bibliography_ROSA}
\appendix
\onecolumngrid
\section{Effect of particle--hole symmetry on thermoelectric operation}
\label{app:phsymmetry}

\subsection{Heat engine operation}
The purpose of this appendix is to show that when the quantum dot level is tuned to the 
Fermi energy, $\epsilon_d = E_F$, the device \emph{cannot} operate as a heat 
engine, meaning it can never produce positive work. Our starting point is to identify two fundamental symmetries  that arise when $\epsilon_d = E_F$. Firstly, the transmission function becomes symmetric 
about the Fermi energy when $\epsilon_d = E_F$, i.e., 
    \[
    \mathcal{T}(E+E_F)\Big\rvert_{\epsilon_d = E_F} 
    = \mathcal{T}(-E+E_F)\Big\rvert_{\epsilon_d = E_F}.
    \]
Physically, electrons at energy $+E$ above and $-E$ below $E_F$ are equally likely to be transmitted. Then, we employ the particle-hole symmetry of the Fermi function, $f(-x) = 1 - f(x)$. This means that the probability of a hole at energy $-x$ equals the  probability of an electron at energy $+x$. Using these two symmetries, the charge current is odd in voltage  $I(V,\theta) = -I(-V,\theta)$. An immediate consequence is that at zero voltage, the current vanishes for \emph{any} thermal bias $\theta$, i.e., $I(V=0, \theta) = 0 \quad \forall\, \theta$ meaning that a temperature difference alone cannot drive a net electrical current through the device,  the Seebeck effect is suppressed, thermopower is absent. Using the
charge conservation law then the currents satisfy $I_L(V,\theta) = -I_L(-V,-\theta)$, resulting into
\[
I(V,\theta) = I(V,-\theta).
\]
This symmetry tells us the current does not depend on the sign of the temperature bias. 
Therefore, without loss of generality, we can restrict our analysis to 
$V,\theta > 0$.

Finally, we now show that 
$I(V>0,\theta>0) \geq 0$. After shifting the integration variable by $E_F$ and 
exploiting the even symmetry of the transmission, the charge current can be 
rewritten as a sum of two manifestly non-negative contributions:

\begin{align*}
I &= \int_{-\infty}^{\infty} dE\, \mathcal{T}(E+E_F)
\biggl[f\Bigl(\frac{E-V/2}{T_L}\Bigr) 
- f\Bigl(\frac{E+V/2}{T_R}\Bigr)\biggr] \\[6pt]
&= \int_{0}^{\infty} dE\, \mathcal{T}(E+E_F)
\biggl[f\Bigl(\frac{E-V/2}{T_L}\Bigr) 
- f\Bigl(\frac{E+V/2}{T_L}\Bigr)\biggr] \\[4pt]
&\quad + \int_{0}^{\infty} dE\, \mathcal{T}(E+E_F)
\biggl[f\Bigl(\frac{E-V/2}{T_R}\Bigr) 
- f\Bigl(\frac{E+V/2}{T_R}\Bigr)\biggr].
\end{align*}

The equality above is obtained by splitting the original integral into 
$E\in(-\infty,0)$ and $E\in[0,\infty)$, and using the two symmetries previously discussed  to fold the negative-energy part onto the positive-energy axis. Each integrand is non-negative since the Fermi function is strictly decreasing, 
and $\frac{E-V/2}{T_{1,2}} \leq \frac{E+V/2}{T_{1,2}}$ for $V>0$ and positive 
temperatures $T_{1,2}$, we have
\[
f\Bigl(\frac{E-V/2}{T_{1,2}}\Bigr) - f\Bigl(\frac{E+V/2}{T_{1,2}}\Bigr) \geq 0.
\]
Since $\mathcal{T}(E) > 0$ for all $E$, both integrals are non-negative, and 
therefore $I(V>0,\theta>0) \geq 0$. The conclusion is that the power output is $\dot{W} = -IV \leq 0$ for 
$V,\theta > 0$, meaning the device always \emph{receives} work rather than 
producing it. Since by symmetry this conclusion holds for all signs of $V$ and 
$\theta$, \emph{the device cannot operate as a heat engine when $\epsilon_d = E_F$}.


\subsection{Refrigerator and heat pump operation}
The goal now is to show that, for the same condition $\epsilon_d = E_F$, 
the device also \emph{cannot} act as a refrigerator or heat pump. 
Refrigeration requires extracting heat from the cold reservoir, i.e.\ 
$J_R \leq 0$ for $\theta\geq 0$ and $J_L \leq 0$ for $\theta\leq 0$. 
We will show both are not possible to achieve.

Following the same symmetry arguments as before, the heat current from 
reservoir R can be split into two convenient pieces:

\begin{align*}
J_R &= \int_{0}^{\infty}dE\, \mathcal{T}(E+E_F)
\Bigl\{(E+V/2)\,f\!\left(\tfrac{E+V/2}{T_R}\right) 
+ (E-V/2)\,f\!\left(\tfrac{E-V/2}{T_R}\right)\Bigr\} \\
&\quad - \int_{0}^{\infty}dE\, \mathcal{T}(E+E_F)
\Bigl\{(E+V/2)\,f\!\left(\tfrac{E-V/2}{T_L}\right) 
+ (E-V/2)\,f\!\left(\tfrac{E+V/2}{T_L}\right)\Bigr\} \\
&\equiv I_a(T_R) - I_b(T_L).
\end{align*}
Now we demonstrate that  $I_a(T) \leq I_b(T)$ by setting both reservoirs at the same temperature, $T_L = T_R = T$, the heat 
current must satisfy $J_R \leq 0$, since in equilibrium no refrigeration can 
occur. Indeed, substituting $T_L = T_R = T$ the expression simplifies to:
\[
J_R\big|_{T_L=T_R=T} 
= V\!\int_{0}^{\infty}dE\,\mathcal{T}(E+E_F)
\left[f\!\left(\tfrac{E+V/2}{T}\right) 
- f\!\left(\tfrac{E-V/2}{T}\right)\right] \leq 0.
\]
This is non-positive because for $V\geq 0$, the integrand is negative 
(the Fermi function is decreasing), while the prefactor $V\geq 0$, so their 
product is $\leq 0$; for $V\leq 0$ the argument is reversed symmetrically. 
This confirms $I_a(T)\leq I_b(T)$ for all $T$.

The key remaining ingredient is to show that $I_b(T)$ is a 
\emph{monotonically increasing} function of $T$, i.e.\ 
$\frac{\partial I_b}{\partial T} \geq 0$. Computing the derivative explicitly:
\[
\frac{\partial I_b(T)}{\partial T} 
= \int_{0}^{\infty}dE\,\mathcal{T}(E+E_F)\,
\frac{-E}{T^2}
\left\{(E+V/2)\,\partial_E f\!\left(\tfrac{E-V/2}{T}\right) 
+ (E-V/2)\,\partial_E f\!\left(\tfrac{E+V/2}{T}\right)\right\},
\]
where we used $\partial_T f\!\left(\frac{E-\mu}{T}\right) 
= -\frac{E-\mu}{T}\,\partial_E f\!\left(\frac{E-\mu}{T}\right)$.

We verify that the integrand is non-negative by splitting the range of 
integration into two sub-intervals:

\begin{itemize}

\item \textbf{Case 1: $0 \leq E \leq V/2$.}
In this range, $E - V/2 \leq 0$, so $\partial_E f\!\left(\frac{E-V/2}{T}\right)$ 
is a negative, decreasing function (since the derivative of the Fermi function 
peaks at zero argument and decays on both sides). Meanwhile, 
$\partial_E f\!\left(\frac{E+V/2}{T}\right)$ is negative and increasing 
for $E \geq 0$. Combining these observations and using 
$\partial_E f \leq 0$ with the monotonicity properties of the Fermi 
derivative, one shows:
\[
(E+V/2)\,\partial_E f\!\left(\tfrac{E-V/2}{T}\right) 
+ (E-V/2)\,\partial_E f\!\left(\tfrac{E+V/2}{T}\right) \leq 0,
\]
so that $\frac{-E}{T^2}$ times this expression is $\geq 0$ for $E\geq 0$.

\item \textbf{Case 2: $E \geq V/2$.}
In this range, both $E + V/2$ and $E - V/2$ are positive.  
Since $\partial_E f < 0$ always:
\[
(E\pm V/2)\,\partial_E f\!\left(\tfrac{E\mp V/2}{T}\right) \leq 0,
\]
so their sum is $\leq 0$, and multiplying by $\frac{-E}{T^2}\geq 0$ 
gives a non-negative contribution.

\end{itemize}

In both cases the integrand is non-negative, confirming:
\[
\frac{\partial I_b(T)}{\partial T} \geq 0 \qquad \forall\, V\geq 0.
\]
Since $I_b$ is monotonically increasing and $T_L \geq T_R$ 
(i.e.\ $\theta \geq 0$), we have $I_b(T_L) \geq I_b(T_R)$. 
Together with $I_a(T_R) \leq I_b(T_R)$ from Step 2, this gives:
\[
J_R = I_a(T_R) - I_b(T_L) \leq I_b(T_R) - I_b(T_L) \leq 0.
\]
Finally, by replacing $V \to -V$ in the expression for $J_R$ and using 
$f(-x) = 1 - f(x)$, one shows $J_R(V,\theta) = J_R(-V,\theta)$, 
so the result holds for \emph{all} values of $V$.

An entirely analogous argument (swapping the roles of reservoirs L and R, 
i.e.\ $T_L \leftrightarrow T_R$) gives $J_L(V,\theta) \leq 0$ for all $\theta \leq 0$. Therefore, regardless of the applied voltage $V$:
\begin{itemize}
    \item $J_R \leq 0$ for all $\theta \geq 0$: the device always 
    releases heat into (rather than extracting it from) the cold reservoir.
    \item $J_L \leq 0$ for all $\theta \leq 0$: the same holds 
    symmetrically for reservoir L.
\end{itemize}
When $\epsilon_d = E_F$, the device cannot operate as 
a refrigerator or heat pump under any combination of voltage and temperature bias.
\section{Charge-current TUR indicator is minimized for symmetric tunnel couplings}
\label{app:tursat}

Here we show that the MTUR is most tightly saturated when $\Gamma_L = \Gamma_R$.
Let us consider two cases, namely (i) the symmetrical case, $\Gamma_L = \Gamma_R$, and the asymmetrical case, $\Gamma_L \neq \Gamma_R$, but both with same mean value $ \Gamma_M := \frac{\Gamma_L + \Gamma_R}{2}$. Note that in the symmetrical case $\Gamma_L = \Gamma_R = \Gamma_M$, and the transmission function is $\mathcal{T}^{(sym)}(E) = \frac{\Gamma_M^2}{(E-\epsilon_d)^2 + \Gamma_M^2}$. Let consider $\eta := \frac{\Gamma_L}{\Gamma_R}$, which is well defined as $\Gamma_L, \Gamma_R \neq 0$. As $\Gamma_L$ and $\Gamma_R$ give two degrees of freedom, the effect of $\Gamma_L$ and $\Gamma_R$ can be completely characterized by $\Gamma_M$ and $\eta$ by performing the variable substitution $(\Gamma_L,\Gamma_R) \to (\Gamma_M,\eta)$ given by

\begin{equation}
\Gamma_L = \frac{2\eta}{\eta+1}\Gamma_M \qquad \Gamma_R = \frac{2}{\eta+1}\Gamma_M
\end{equation}

Then, the transmission function can be expressed as

\begin{equation}
    \mathcal{T}(E) = \frac{4\eta}{(\eta+1)^2}\frac{\Gamma_M^2}{(E-\epsilon_d)^2 + \Gamma_M^2} = \frac{4\eta}{(\eta+1)^2}\mathcal{T}^{(sym)}(E)
    \label{tasym}
\end{equation}

Then, the currents and entropy production rate can be expressed as

\begin{equation}
    I_\alpha = \frac{4\eta}{(\eta+1)^2} I_\alpha^{(sym)} \qquad \qquad J_\alpha = \frac{4\eta}{(\eta+1)^2} J_\alpha^{(sym)} \qquad \qquad \dot{\sigma} = \frac{4\eta}{(\eta+1)^2} \dot{\sigma}^{(sym)}
\end{equation}
 
where the symmetrical quantities go with symmetrical transmission $\mathcal{T}^{(sym)}(E)$ in the integral. 

On the other side, the noise for the current $I_\alpha$ can be expressed as sum of two contributions one which goes with $\mathcal{T}(E)$, and the other which goes with $\mathcal{T}^2(E)$.
\begin{gather*}
   S_\alpha =\int_{-\infty}^{\infty} dE \mathcal{T}(E) \left[ f_L(E)(1 - f_L(E)) + f_R(E)(1 - f_R(E)) + (f_L(E) - f_R(E))^2 \right] - \\
   \int_{-\infty}^{\infty} dE \mathcal{T}^2(E)(f_L(E) - f_R(E))^2 = S_{\alpha, \,a}-S_{\alpha, \, b}
    \label{noiseasymetry}
\end{gather*}
where $S_{\alpha,a} \equiv \int dE\, T(E)\!\left[f_L(1-f_L) + f_R(1-f_R) + (f_L-f_R)^2\right]$ and $S_{\alpha,b} \equiv \int dE\, T^2(E)(f_L-f_R)^2$ denote the $T$ and $T^2$ weighted noise contributions, respectively. Making the substitution of $\mathcal{T}(E)$, then $S_{\alpha, \,a} = \frac{4\eta}{(\eta+1)^2} S_{\alpha,\, a}^{(sym)}$, since it goes with $\mathcal{T}(E)$, and $S_{\alpha, \, b} = \frac{16\eta^2}{(\eta+1)^4}$, since it goes with $\mathcal{T}^2(E)$. Therefore, we obtain
\begin{gather*}
  S_\alpha = S_{\alpha, \,a}-S_{\alpha, \, b} = \frac{4\eta}{(\eta+1)^2} S_{\alpha,\, a}^{(sym)} - \\
  \frac{16\eta^2}{(\eta+1)^4} S_{\alpha, \, b}^{(sym)} = \frac{4\eta}{(\eta+1)^2} S_{\alpha,\, a}^{(sym)} - \frac{4\eta}{(\eta+1)^2}S_{\alpha, \, b}^{(sym)} + \frac{4\eta}{(\eta+1)^2}S_{\alpha, \, b}^{(sym)}-\frac{16\eta^2}{(\eta+1)^4} S_{\alpha, \, b}^{(sym)} = \\\frac{4\eta}{(\eta+1)^2} \left\{(S_{\alpha,\, a}^{(sym)} - S_{\alpha, \, b}^{(sym)}) + \left(1-\frac{4\eta}{(\eta+1)^2}\right)S_{\alpha, \, b}^{(sym)}\right\}= \frac{4\eta}{(\eta+1)^2} \left\{S_\alpha^{(sym)} + \frac{(\eta-1)^2}{(\eta+1)^2}S_{\alpha, \, b}^{(sym)}\right\}
\end{gather*}

\begin{equation}
   \implies  S_\alpha = \frac{4\eta}{(\eta+1)^2} \left\{S_\alpha^{(sym)} + \frac{(\eta-1)^2}{(\eta+1)^2}S_{\alpha, \, b}^{(sym)}\right\}
\end{equation}

Introducing all terms in TUR coefficient, we obtain 

\begin{gather*}
    F = \frac{\sigma S}{I^2} -2 =  \frac{\cancel{\frac{4\eta}{(\eta+1)^2}}\sigma^{(sym)} \cancel{\frac{4\eta}{(\eta+1)^2}} \left\{S^{(sym)} + \frac{(\eta-1)^2}{(\eta+1)^2}S_{b}^{(sym)}\right\}}{\cancel{ \left(\frac{4\eta}{(\eta+1)^2} \right)^2 }{I^{(sym)}}^2} -2 =\\
    \underbrace{\frac{\sigma^{(sym)}S^{(sym)}}{{I^{(sym)}}^2}-2}_{= \mathcal{F}^{(sym)}}+\frac{(\eta-1)^2}{(\eta+1)^2}\frac{\sigma^{(sym)}S^{(sym)}_b}{{I^{(sym)}}^2} =
\mathcal{F}^{(sym)}+ \frac{(\eta-1)^2}{(\eta+1)^2}\frac{\sigma^{(sym)}S^{(sym)}_b}{{I^{(sym)}}^2} = \\
    \mathcal{F}^{(sym)} + \frac{(\eta-1)^2}{(\eta+1)^2}\frac{\sigma^{(sym)}}{{I^{(sym)}}^2}    \int_{-\infty}^{\infty} dE {\mathcal{T}^{(sym)}}^2(E)\left(f_L(E) - f_R(E)\right)^2
\end{gather*}

\begin{equation}
 \implies    \Delta \mathcal{F} = \mathcal{F} - \mathcal{F}^{(sym)} = \frac{(\eta-1)^2}{(\eta+1)^2}\frac{\sigma^{(sym)}}{{I^{(sym)}}^2} \int_{-\infty}^{\infty} dE {\mathcal{T}^{(sym)}}^2(E)\left(f_L(E) - f_R(E)\right)^2 \geq 0 \, \,
    \label{dF}
\end{equation}

This difference is always positive, since $\sigma \geq 0$ and all the other terms are quadratic. Therefore for a given and arbitrary $\Gamma_L$ and $\Gamma_R$, there always exists another configuration of these with the same mean value $\Gamma_M$, which is a symmetrical configuration $\Gamma_L = \Gamma_R$, where $\mathcal{F}$ is smaller. In other words, for the same mean value $\Gamma_M$, the symmetrical case is always optimal.
\section{Analytic expression for thermal noise}
\label{app:thermalnoise}
The charge and heat currents can be solved analytically by expressing de Fermi functions in terms of the digamma functions $\Psi$ and using residue theorem to integrate over a semi disk contour \cite{PhysRevB.91.165431}. 
Using the relation $f_\alpha(1-f_\alpha) = - k_B T_\alpha \, \partial f_\alpha / \partial E$, the thermal contribution to the charge-current autocorrelation in a single reservoir, $S_I^{\rm th}$, can be expressed in terms of trigamma functions $\Psi^{(1)}$ as
\begin{equation}
        S_I^{th} =
    \frac{e^2}{h} \frac{\Gamma_L\Gamma_R}{4\pi^2}\sum_\alpha\int_{-\infty}^{\infty} dE \left\{ \frac{1}{(E - \epsilon_d + i \Gamma)(E - \epsilon_d - i \Gamma)} \left [ \Psi^{(1)}(s_\alpha^+)  + \Psi^{(1)}(s_\alpha^-) \right ]\right\}
\end{equation}
with $s_\alpha^\pm = \frac{1}{2} \pm i \frac{E - \mu_\alpha}{2\pi k_B T_\alpha}$. This integral can be also solved analytically using residue theorem integrating over a semi disk contour, and, as $\Psi^{(1)} (z^*) = \left[ \Psi^{(1)} (z) \right]^*$, we obtain

\begin{equation}
    S_I^{th} = \frac{e^2}{h} \frac{\Gamma_L\Gamma_R}{\pi(\Gamma_L + \Gamma_R)}\sum_\alpha  \operatorname{Re}\left[  \Psi^{(1)}\left(\frac{1}{2} + i \frac{\epsilon_d-\mu_\alpha-i\frac{\Gamma_L+\Gamma_R}{2}}{2 \pi k_B T_\alpha}\right) \right]
\end{equation}

Analogous analytic expressions for the heat-current autocorrelation $S^{\rm th}_J$ and the mixed charge–heat correlator $S^{\rm th}_{IJ}$ can be derived by the same method (replacing the prefactor $X(\epsilon)$ accordingly). For brevity, these contributions are evaluated numerically in the present work.
\end{document}